\pdfoutput=1
\documentclass[twocolumn,showpacs,linenumbers,reprint, graphicx,amsmath,amssymb]{revtex4}

\usepackage{graphicx}
\usepackage{dcolumn}
\usepackage{bm}
\usepackage[pdftex]{hyperref}
\hypersetup{
    colorlinks,%
    citecolor=blue,%
    filecolor=black,%
    linkcolor=black,%
    urlcolor=blue
}

\begin{document}


\title{Effect of high-$\kappa$ gate dielectrics on charge transport in graphene-based field effect transistors (g-FET)}
\author{Aniruddha Konar}
\email[{\bf{E-mail:}}]{akonar@nd.edu} 
\author{Tian Fang}
\author{ Debdeep Jena}
\affiliation{Department of Physics and Department of Electrical Engineering , University of Notre Dame, Notre Dame, Indiana, IN 46556, USA.}
\date{\today}

\begin{abstract}
The effect of various dielectrics on charge mobility in single layer graphene is investigated.  By calculating the remote optical phonon scattering arising from the polar substrates, and combining it with their effect on Coulombic impurity scattering, a comprehensive picture of the effect of dielectrics on charge transport in graphene emerges.  It is found that though high-$\kappa$ dielectrics can strongly reduce Coulombic scattering by dielectric screening, scattering from surface phonon modes arising from them wash out this advantage. Calculation shows that within the available choice of dielectrics, there is not much room for improving carrier mobility in actual devices at room temperatures.
\end{abstract}

\pacs{72.b-g}
\maketitle
\section{Introduction}

Graphene, a 2D gapless semiconductor with a honeycomb crystal structure of carbon atoms, has gained significant attention recently owing to its conical bandstructure \cite{Novoselovnature05}, unconventional non-integer quantum Hall effect \cite{zhangnature05,zhangscience07}, possible applications emerging from high room-temperature (RT) electron mobility ($\thicksim 10^4$ cm$^{2}$/V$\cdot$s) \cite{novoselovscience04} ,and tunable bandgaps in nanostructures carved from it \cite{HanPRL07, HdaiPRL08}.  Appreciable modulation of current by electrostatic gating indicate graphene as an excellent material for conventional and possibly novel electronic devices \cite{MericNatNano08}.  High mobilities can facilitate coherent ballistic transport over large length scales.  This can enable novel devices that exploit the wavelike features of electrons, resulting in `electron optics' in the nanoscale material \cite{cheianovScience07}.  

For graphene located in close proximity to dielectric substrates, the highest RT carrier mobilities experimentally reported are in the $\thicksim 10^4$ cm$^{2}$/V$\cdot$s range \cite{ChenNatNano08} for 2D carrier concentrations in the $\sim 10^{12}$/cm$^2$ regime.  However, suspending graphene by removing the underlying substrate and driving off impurities sticking to it resulted in much higher mobilities near RT, albeit at low carrier concentrations \cite{BolotinPRL08, BolotinSSC08, andreiNatNano08}.  When compared to 2D electron gases (2DEGs) in narrow-bandgap III-V semiconductor heterostructures such as InAs and InSb \cite{bennettSSE05}, mobilities and low-field conductivities (charge $\times$ mobility) in 2D graphene on dielectric substrates reported to date are lower.  

Recently, it was predicted that for thin semiconducting nanostructures, impurity scattering can be reduced by surrounding the structure with high-$\kappa$ dielectrics \cite{JenaPRL07, KonarJap07}, and a recent experimental observation was made for graphene \cite{jangPRL08}.  The reduction of impurity scattering is affected by the reduction of Coulomb scattering by the high-$\kappa$ dielectric.  However, several recent works have pointed out that high-$\kappa$ dielectrics in close proximity with a conducting channel in a semiconductor lead to enhanced surface-optical (SO) phonon scattering due to remote optical phonon coupling between electrons in the channel and polar vibrations in the dielectric.  This feature presents itself not only in graphene \cite{FratiniPRB08, ChenNatNano08, ponomarenkoPRL}, but in carbon nanotubes \cite{perebeinosNanoLett08}, as well as in the workhorse of the electronics industry - Silicon Metal-Oxide Field-Effect Transistors (MOSFETs) \cite{FischettiJAP01}.  

At low electric fields, intra- and intervalley optical phonon scattering is negligible due to the insufficient kinetic energy of carriers.  If a nominal charged impurity concentration of $n_{imp} \sim 10^{11}$/cm$^2$ is present at a graphene/high-$\kappa$ dielectric interface, for carrier concentrations $\leq 10^{12}$/cm$^2$, acoustic phonon scattering at RT is also relatively insignificant owing to the low density of states of graphene at the Fermi energy (see  \cite{hwangacPRB07, VaskoPRB07, McdonaldPRL07, hwangPRL07}).  In this technologically relevant regime, the competing effects of impurity and SO-phonon scattering are responsible for the low-field transport properties of graphene.  High-$\kappa$ gate oxides result in good electrostatic gate control of the Fermi level in the graphene layer, and at the same time reduce Coulombic scattering from charged impurities.  However, at the same time, they give rise to SO phonon scattering.  It is imperative at this stage that this twofold role of dielectrics be studied in detail to clarify the competing roles, and consequently outline routes for retaining the high mobility and mean free path of carriers in graphene.  That is the goal of this work.
\section{Surface Phonon at Graphene/insulator Interface}
\subsection{Dielectric continuum model of surface phonons}
Surface modes in the phonon spectrum of a solid arise due to its finite size.  In the case of a polar material, the phonon field due to longitudinal surface modes propagates along the surface and induces a non-vanishing decaying electric field outside it \cite{Aschcroft}.  The dispersion of these surface optical (SO) phonons and the induced electric field in the inversion layer of a semiconductor-oxide interface was first calculated by Wang and Mahan \cite{MahanPRB72}.  Later, Fischetti {\em et. al}  \cite{FischettiJAP01} solved the dynamical dielectric response of a coupled channel/insulator/gate system and showed that the channel and gate plasmons can alter the dispersion relation of SO-phonon modes at semiconductor-oxide interface for a Si-MOSFET structure.  In this work, we consider the gate as an ideal metal, implying that all electric field lines originating from time-dependent Coulombic fluctuations at the gate/oxide interface should terminate on the gate metal \cite{KotlyarIEDM04} producing an insignificant effect on the channel (in our case  graphene) dielectric response.
\begin{figure}
\includegraphics[width=70mm]{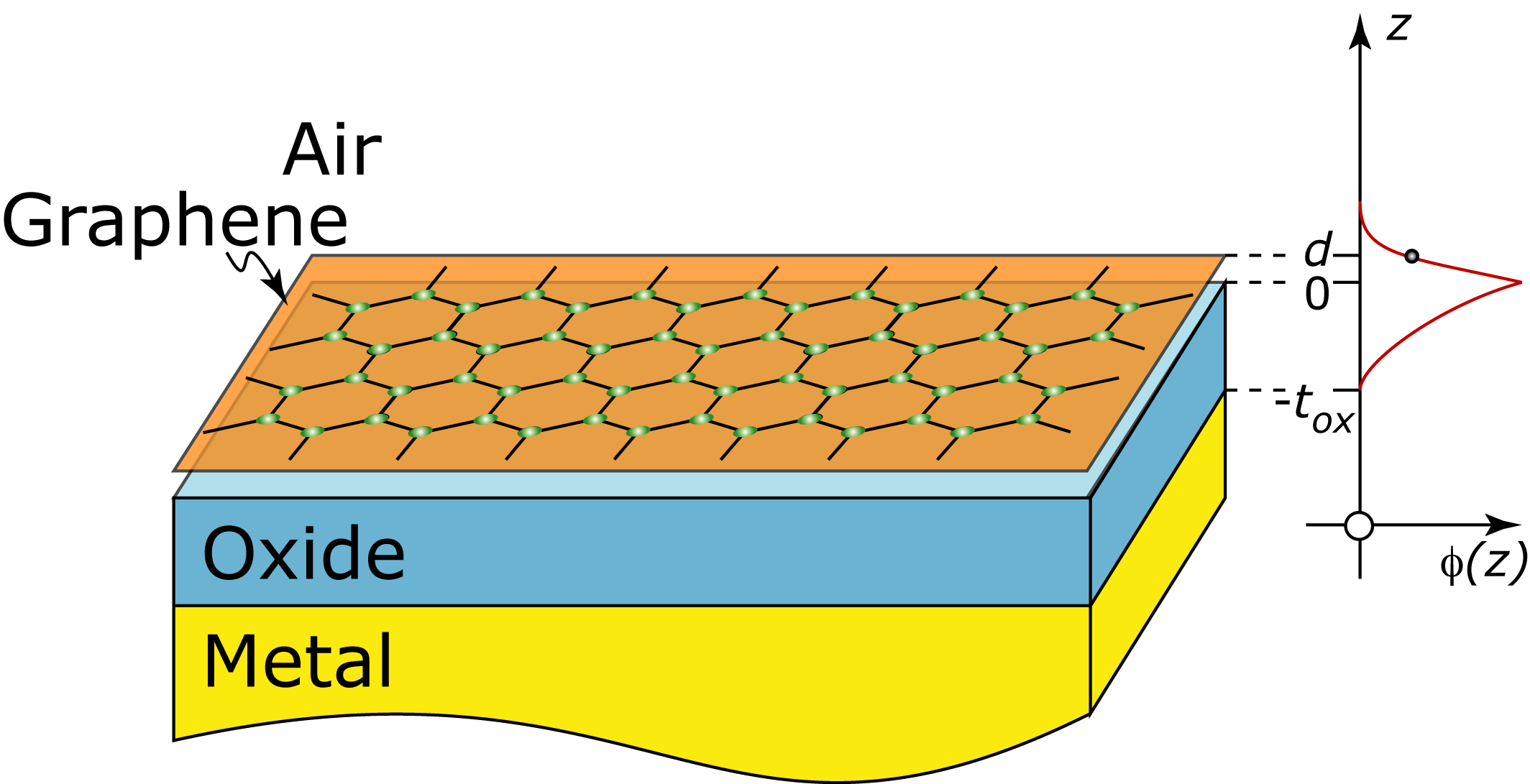}
\caption{Graphene/Oxide/Metal layer structure, and a schematic sketch of the surface-optical (SO) phonon mode strength.}
\label{Fig1}
\end{figure}
Consider the structure as shown in Fig \ref{Fig1}.  A dielectric of thickness $t_{ox}$ is deposited on a metal gate, and a single layer graphene is placed at an equilibrium distance $d$ from the oxide, and air ($\kappa$ = 1) covers the rest of the space.  Denoting ${\bf{q}}$ and ${\bf{r}}$ as the two-dimensional wave vector and spatial vector in the graphene plane, we write the time-dependent SO-phonon field at a point $({\bf{r}},z)$ as
\begin{eqnarray}
\phi({\bf{r}},z,t)=\sum_{{\bf{q}}}\phi_{q,\omega}(z)e^{i({\bf{q}}.{\bf{r}}-\omega t)},
\end{eqnarray}
where the components of SO-modes $\phi_{q,\omega}(z)$ are solutions of Maxwell's equation given by \cite{FischettiJAP01}
\begin{eqnarray}
\phi_{q,\omega}(z)=\begin{cases}
0, & -\infty\leq z \leq -t_{ox} \\
2b_{q,\omega} \sinh q(z+t_{ox}),    & -t_{ox}\leq z \leq 0 \\
d_{q,\omega}e^{-qz},  & z\ge 0.
\end{cases}
\label{potential}
\end{eqnarray}
Here, $b_{q,\omega}$ and $d_{q,\omega}$  are normalization constants.  Applying electrostatic boundary conditions at the interface, we get the secular equation

\begin{eqnarray}
\epsilon_{ox}(\omega)+\epsilon_{s}(q,\omega)\tanh(qt_{ox})=0,
\label{secularequation}
\end{eqnarray}
where, $\epsilon_{ox}(\omega)$ and $\epsilon_{s}(q,\omega)$ are the dynamic dielectric functions of the oxide and graphene  respectively. For bulk dielectrics, the frequency dependent dielectric function can be written as 
\begin{figure}
\includegraphics[width=85mm]{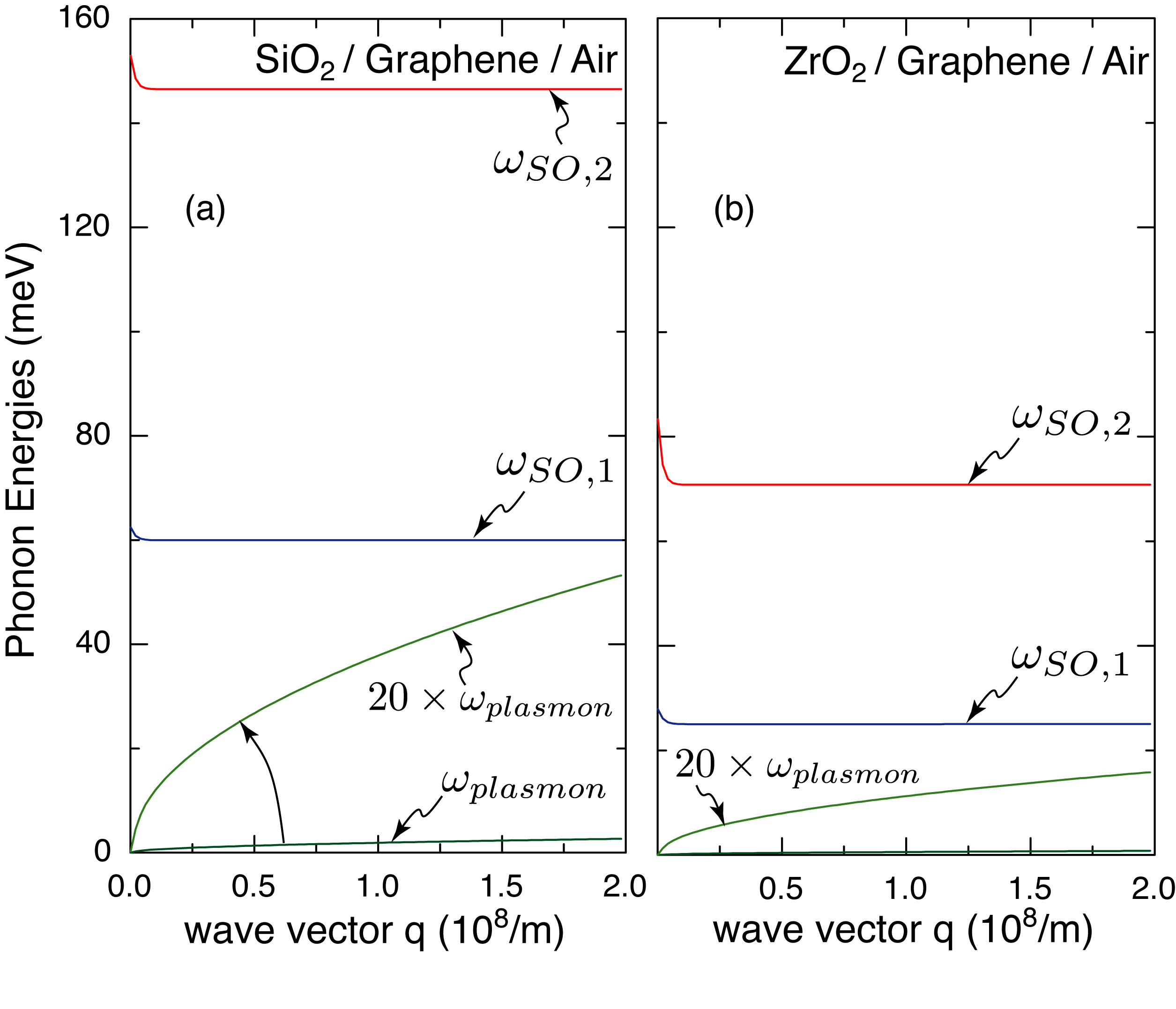}
\caption{Energy dispersion $\omega(q)$ of surface-optical (SO) modes for Graphene on SiO$_{2}$ and ZrO$_2$.  The two SO phonon modes labeled $\omega_{SO,1},\omega_{SO,2}$ are nearly dispersionless (except near $q \approx 0$), and their magnitudes are close to the values in the absence of graphene.  The graphene plasmon mode is energetically lower than surface modes.}
\label{Fig2}
\end{figure} 
\begin{eqnarray}
\epsilon_{ox}(\omega)=\epsilon_{\infty} + \sum_{n} \frac{ f_{n} (\omega_{TO}^{n})^{2} }{(\omega_{TO}^{n})^{2} -\omega^{2}}.
\end{eqnarray}
Here, $\epsilon_{\infty}$ is the high frequency ($\omega \to \infty$) permittivity, $\omega_{TO}^{n}$ is the $n^{th}$ bulk transverse optical phonon frequency and $f_{n}$ is the corresponding oscillator strength of the dielectric oxide. The values of oscillator strength for different dielectrics are extracted from experimental data as outlined in ref. \cite{FischettiJAP01}; the material constants are listed in Table \ref{table1}.  
In the long-wavelength limit ($q\to 0$), the dielectric function of graphene in the RPA approximation is \cite{HwangPRB07}
\begin{eqnarray}
\epsilon_{s}(q,\omega)=\epsilon_{s}^{\infty}\Big(1-\frac{\omega_{p}^{2}(q)}{\omega^{2}}\Big).
\label{plasmon}
\end{eqnarray}
Here, $\epsilon_{s}^{\infty}=(\epsilon_{ox}^{\infty}+1)/2$, $\omega_{p}(q)=  v_{F}\sqrt{ \alpha q \sqrt{\pi n}}$ is the plasma frequency of single layer graphene, $\alpha = e^{2}/2 \pi \epsilon_{0} \kappa_{av} \hbar v_{F}$ is the effective fine structure constant of graphene, $e$ is the electron charge, $\epsilon_{0}$ is the permittivity of vacuum, $\hbar = h/2\pi$ is the reduced Planck's constant, $v_{F} \sim 10^{6}$ m/s is the Fermi velocity characterizing the bandstructure of graphene, $n$ is the 2D carrier density, and $\kappa_{avg}$ is the average dielectric constant of the system.  For example, for graphene sandwiched between SiO$_{2}$ ($\kappa_{si}=3.9$) and air ($\kappa=1$), $\kappa_{avg}=(3.9+1)/2 = 2.45$.  The numerical solution of Eq. \ref{secularequation} has three roots.  Two of them represent interfacial phonon modes corresponding to two bulk TO modes of the dielectric, and the third root represents the graphene plasmon mode.  Figs \ref{Fig2} (a) and (b) show the calculated dispersions of these three modes for a graphene layer on SiO$_{2}$ and ZrO$_{2}$.  The SO phonon modes are found to be nearly dispersionless, except at the lowest wavevectors.  The calculation of these dispersions show that the SO-phonon modes of the oxide-graphene-air system are not too different from the oxide-air surface.  The effect of the graphene dielectric response on the SO modes is negligible, since charges in graphene merely screen the electric field lines originating from the oxide-graphene interface.  From the results in Fig \ref{Fig2}, we observe that $\omega_{plasmon} << \omega_{SO,1},\omega_{SO,2}$.  Hence, for the rest of the work we use the calculated dispersionless part of the SO phonon modes - the energies of these two modes for some of the commonly used dielectric gates are given in Table \ref{Table1}.
\begin{table}
\caption{\label{table1}Surface-optical phonon modes for different dielectric gates. Parameters have been taken from ref. \cite{FischettiJAP01}.}
\begin{ruledtabular}
\begin{tabular}{ccccccc}
					&  SiO$_{2}$	&AlN		&Al$_{2}$O$_{3}$	&HfO$_{2}$ 	&  ZrO$_{2}$	& SiC\footnotemark[1] 	\\
\hline
$\kappa_{ox}^{0}$   		&  3.9  		&  9.14  	&  12.53  			&  22.0  		&  24.0 		& 9.7					\\
$\kappa_{ox}^{\infty}$  	&  2.5  		&  4.8  	&  3.2  			&  5.03  		&  4.0 		& 6.5					\\ 
$\omega_{TO,1}$   		&  55.6  		&  81.4 	&  48.18  			&  12.4  		&  16.67		&  -					\\
$\omega_{TO,2}$   		&  138.1		&  88.5  	&  71.41  			&  48.35  		&  57.7		&  -					\\
$\omega_{SO,1}$  		&  59.98  		&  83.60  	&  55.01  			&  19.42  		&  25.02 		&  116\footnotemark[2]	\\ 
$\omega_{SO,2}$  		&  146.51  	&  104.96 	&  94.29 			&  52.87 		&  70.80 		&  167.58				\\
\end{tabular}
\end{ruledtabular}%
\footnotetext[1]{Refererence \cite{FratiniPRB08}.}
\footnotetext[2]{Single surface phonon mode measured in reference \cite{NienhausSurfSci89}.}
\label{Table1}
\end{table}
\subsection{Interaction Hamiltonian and scattering rates}
With the calculated SO-phonon frequencies above, the Hamiltonian of the electron-SO phonon system is given by

\begin{equation}
\mathcal{H}=\sum_{k}\mathcal{E}_{k}c^{\dagger}_{k}c_{k}+\sum_{q,\nu}\hbar\omega_{so}^{\nu}a^{\nu\dagger}_{q}a^{\nu}_{q}+H_{int},
\label{ham}
\end{equation}

where $\mathcal{E}(k)= \pm\hbar v_{F}|k|$ is the kinetic energy of massless Dirac fermions close to the ($\mathcal{K, K'}$) points of the graphene Brillouin Zone, and $c^{\dagger}_{k}(c_{k})$ is the electron creation (annihilation) operator with wavevector $k$.  The second term of Eq. (\ref{ham}) is the phonon part of the total Hamiltonian, where $a^{\nu\dagger}_{q}(a^{\nu}_{q})$ represents the surface phonon creation (annihilation) operator for the $\nu^{th}$ mode. The electron-phonon interaction part of the Hamiltonian is given by 

\begin{equation}
H_{int} = e \mathcal{F}_{\nu} \sum_{q} \Big[\frac{e^{-qz}}{\sqrt{q}}(e^{i{\bf{q}}.{\bf{r}}}a^{\nu\dagger}_{q}+e^{-i{\bf{q}}.{\bf{r}}}a^{\nu}_{q})\Big],
\label{interaction}
\end{equation}

where the electron-phonon coupling parameter $\mathcal{F}_{\nu}$ is given by \cite{MahanPRB72,SakPRB72, MoriPRB89} (also see appendix B)

\begin{equation}
\mathcal{F}_{\nu}^{2} = \frac{\hbar\omega_{SO}^{\nu}}{2 A \epsilon_{0}}\Big(\frac{1}{\kappa_{ox}^{\infty}+1}-\frac{1}{\kappa_{ox}^{0}+1}\Big),
\label{F}
\end{equation}

where, $\kappa_{ox}^{\infty} (\kappa_{ox}^{0})$ is the high (low) frequency dielectric constant of the dielectric, and $A$ is the area of graphene.  We treat this interaction term perturbatively to calculate the electron-SO phonon scattering rates.  The wavefunction for the coupled electron-phonon system can be written as $|\Psi\rangle=|\mathcal{\chi}_{e}\rangle|0\rangle$, where $|0\rangle$ is the phonon vacuum state and $|\mathcal{\chi}_{e}\rangle$ is the electron wavefunction of graphene close to the Dirac points.  The real-space representation of the 2D electron wavefunction of graphene near the ($\mathcal{K, K'}$) points is given by the spinor form

\begin{equation}
\chi_{e}( {\bf k,r} ) = \langle{\bf{r}}|\mathcal{\chi}_{e}\rangle = \frac{ e^{ i {\bf{k}} \cdot {\bf{r}} } }{ \sqrt{2A} } \left( \begin{array}{c} e^{-i\phi} \\1\end{array}\right),
\end{equation}

where $\phi=\tan^{-1}(k_{y}/k_{x})$, and ${\bf k, r}$ are the wavevector and spatial coordinate in the graphene plane.  Since carriers are confined to the graphene plane, the $z-$extent of the electron wavefunction is assumed to be of the form $|\langle{ z} | \mathcal{\chi}_{e} \rangle |^{2} = \delta(z-d) $,  $d$ is the equilibrium distance of the graphene sheet from the oxide surface.  In one-phonon  scattering process, the scattering matrix element is 

\begin{eqnarray}
|\mathcal{M}_{q}^{\nu}(q,\theta)|^{2} & = & |\langle\Psi_{k_{f}}|H_{int}|\Psi_{k_{i}}\rangle|^{2} \nonumber \\
                                                                 & = & e^{2} \mathcal{F}_{\nu}^{2}\Big(\frac{e^{-2qd}}{q}\Big)\cos^{2}\Big(\frac{\theta}{2}\Big),
 \label{matrix}
\end{eqnarray}

where, $q=| {\bf k_{f}} - \bf{k_{i}} |$ is the difference between the initial and final momentum state of electron and $\theta$ is the scattering angle. 

\begin{figure}
\includegraphics[width=85mm]{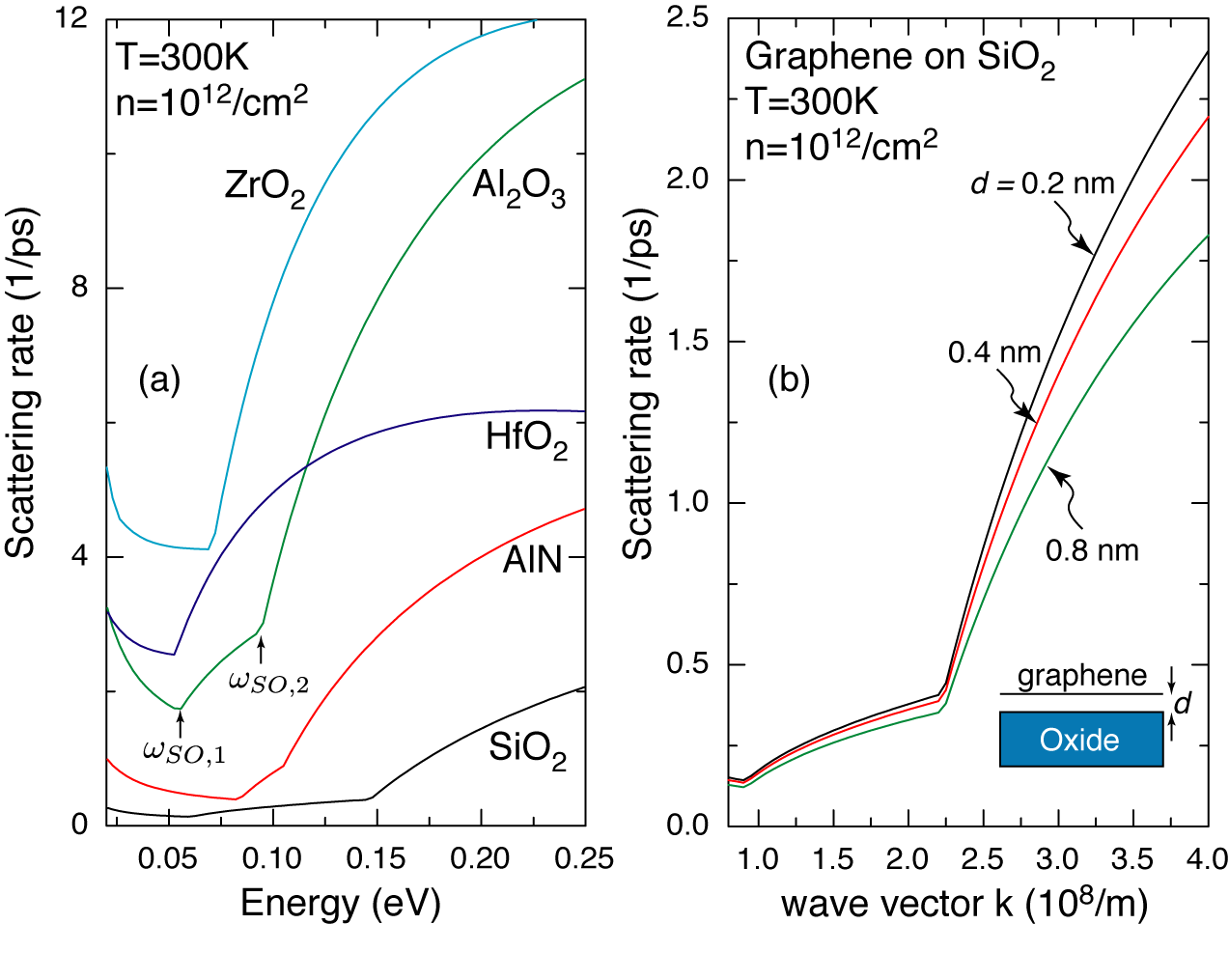}
\caption{Left - (a): SO phonon absorption + emission scattering rates for graphene on various dielectrics.  The higher the static dielectric constants, the higher the SO phonon scattering rate.  The onsets of SO phonon emission into two modes are indicated by arrows.  Right - (b): SO phonon scattering rate for graphene on SiO$_2$ for three different graphene-oxide thicknesses.}
\label{Fig3}
\end{figure}

The electron-SO phonon scattering rate in the relaxation time approximation \cite{Seeger, Chattopadhaya} is then given by

\begin{eqnarray}
\frac{1}{\tau_{SO,\nu}(k,T)}&&=\frac{2\pi}{\hbar}\sum_{q}\Big|\frac{\mathcal{M}_{\nu}}{\epsilon_{2D}(q,0)}\Big|^{2}\Big(1-\frac{ {\bf k_{f} \cdot k_{i}} }{ |{\bf k_{i}}| |{\bf k_{f}}| }\Big)\nonumber\\
&&\times\Big[n_{q,\nu}\delta(\mathcal{E}_{k_{f}}-\mathcal{E}_{k_{i}}-\hbar\omega_{so}^{\nu})+\nonumber\\&&
(1+n_{q,\nu})\delta(\mathcal{E}_{k_{f}}-\mathcal{E}_{k_{i}}+\hbar\omega_{so}^{\nu})\Big],
\end{eqnarray}

where $n_{q,\nu} = 1/( \exp(\hbar\omega_{q,\nu}/kT)  - 1) $ is the equilibrium phonon occupation number and $\epsilon_{2D}(q,0)$ is Thomas-Fermi screening factor of the 2D carriers \cite{Davies}.  Defining dimensionless variables $ u = d q_{0, \nu} $ and $ x = k / q_{0, \nu} $, where $ q_{0,\nu} = \omega_{q,\nu} / v_{F}$ is a constant surface-phonon wave-vector, the scattering rate can be written in a more compact form as

\begin{eqnarray}
\frac{1}{\tau^{\pm}_{s,\nu}(x)} = \frac{ n_{q,\nu}^{\pm} e^{2} \mathcal{F}^{2}_{\nu} A }{ 2 \pi \hbar^{2} v_{F} } \Big(\frac{x+1}{x}\Big) \times I^{\pm}(x,u),
\end{eqnarray}

where, $I^{\pm}(x,u)$ is a dimensionless integral that can be evaluated numerically \cite{int}, and $\pm$ stands for phonon emission ($n^{+}_{q,\nu}=1+n_{q,\nu}$) and absorption ($n^{-}_{q,\nu}=n_{q,\nu}$) respectively.  The total scattering rate is obtained by summing the emission + absorption processes over all SO phonon modes -
\begin{equation}
\frac{1}{\tau_{ph}(k,T)}=\sum_{\nu}\frac{ \Theta [ \mathcal{E}(k) - \hbar \omega_{s,\nu} ] }{\tau_{s,\nu}^{+}(k,T)}+\frac{1}{\tau_{s,\nu}^{-}(k,T)},
\end{equation}
where $ \Theta [ ... ] $ is the Heaviside unit-step function.  Figure \ref{Fig3}(a) shows the SO phonon scattering rates at room temperature calculated using the above formalism for graphene located on five different polar gate dielectrics.  For this plot, a 2D carrier concentration of $n = 10^{12}$/cm$^{2}$ and an equilibrium graphene-oxide distance of $d=0.4$ nm is used \cite{FratiniPRB08}.  Figure \ref{Fig3}(b) shows the dependence of the SO phonon scattering rate on the graphene-dielectric distance for three different values of $d$.  As expected from the decay of the SO phonon evanescent mode from the dielectric surface, the scattering rate reduces with increasing $d$.  The kinks in the scattering rates in Fig \ref{Fig3}(a) indicate the onsets of SO phonon emission processes, as indicated by arrows.  ZrO$_{2}$ with the highest low-frequency dielectric constant ($\kappa^{0} \sim 24$) shows the strongest electron-SO phonon scattering among the five, and SiO$_{2}$ shows the lowest.  If SO phonons were the sole scattering mechanism responsible for limiting the low-field mobility of graphene, this analysis would indicate that using a low-$\kappa$ dielectric would be the most beneficial.  However, the presence of charged impurities and Coulombic scattering due to them changes this simplistic picture.

\subsection{Ionized Impurity Scattering}
The scattering rate due to charged impurities present in the graphene-dielectric interface is given by 
\begin{equation}
\frac{1}{\tau_{imp}(k)}=\frac{n_{imp}}{\pi\hbar}(\frac{e^{2}} {2\epsilon_{0}\kappa_{avg}})^{2}\frac{F(a)}{\mathcal{E}(k)} = \pi \alpha^{2} F(a) \frac{n_{imp} v_{F}}{k},
\label{impscat}
\end{equation}
where $n_{imp}$ is the sheet density of impurities at the interface, $a = \alpha (k_{F}/k)$ is a dimensionless argument with $\alpha$ the effective fine-structure constant defined earlier following Eq. \ref{plasmon}, $k_{F}$ is the Fermi wavevector, and the dimensionless function $F(a)$ is defined in ref. \cite{hwangPRL07}.  $\kappa_{avg}$ is given by the average relative dielectric constant of the two regions surrounding graphene: $\kappa_{avg} = (\kappa_{top} + \kappa_{bottom})/2$.  In typical cases, it is SiO$_{2}$ and air, implying $\kappa_{avg} = (1 + \kappa_{SiO_{2}})/2 \approx 2.45 $. 

\begin{figure}[b]
\includegraphics[width=85mm]{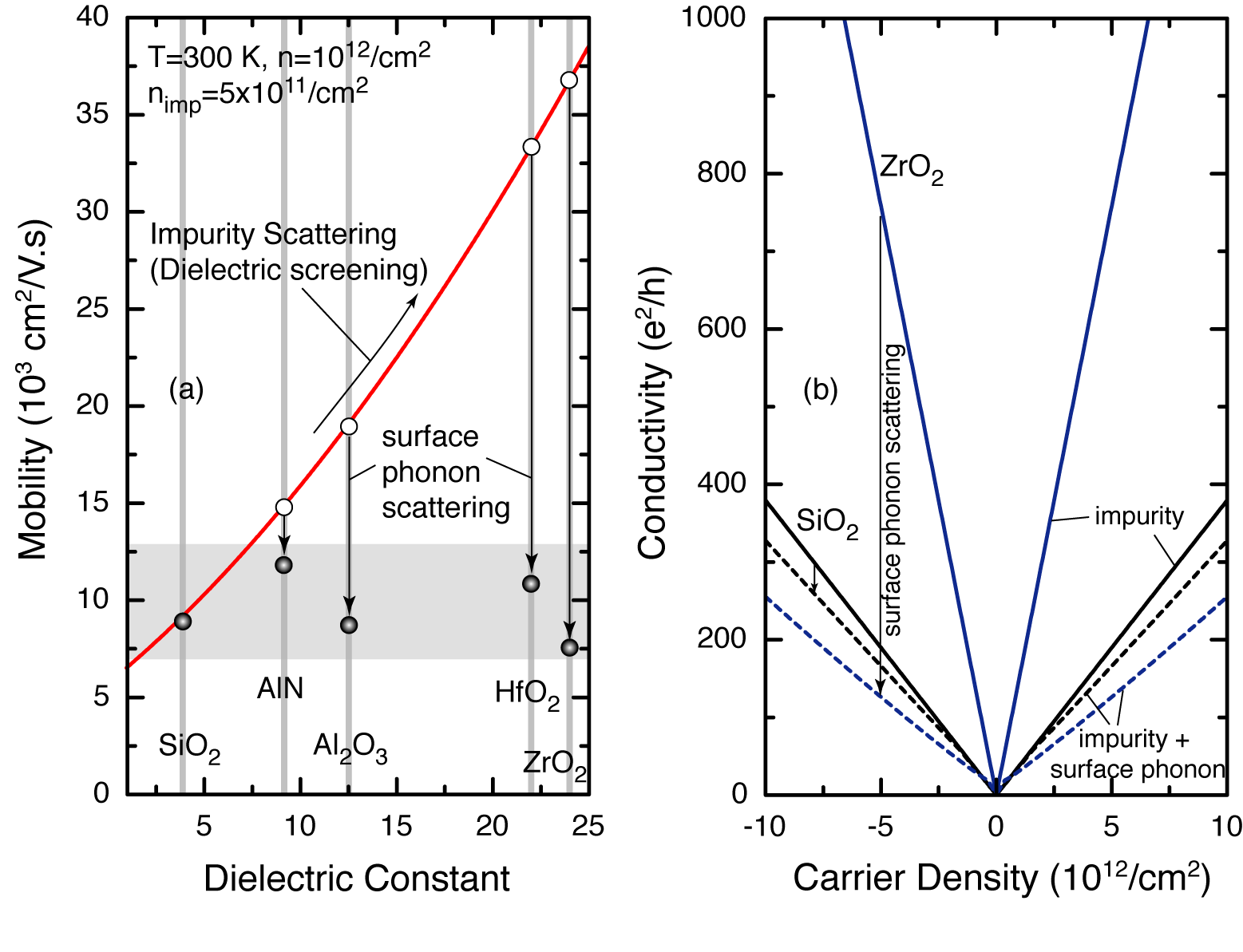}
\caption{ (Left - (a): Electron mobility in graphene as a function of the gate dielectric constant.  High-$\kappa$ dielectrics reduce Coulombic impurity scattering, but strong SO phonon scattering by them reduces the RT mobilities to $\sim 10^{4}$ cm$^2$/V$\cdot$s.  Right - (b): Electron and hole conductivity as a function of carrier concentration for graphene on SiO$_2$ and ZrO$_2$.}
\label{Fig4}
\end{figure} 
\section{Results and Discussions}
As it is clear from Eq. \ref{impscat},  the impurity scattering rate can be strongly suppressed by using high-$\kappa$ dielectric materials next to graphene.  This was pointed out earlier in \cite{JenaPRL07} for the general case of semiconductor nanoscale membranes.  However, using a high-$\kappa$ material also enhances surface-optical phonon scattering rates, as based on the earlier analysis in this work.  With increasing dielectric constant of the graphene environment, impurity scattering is damped, but SO phonon scattering increases, indicating an optimal choice of the dielectric exists for obtaining the highest mobilities for graphene located on substrates. Figure \ref{Fig4}(a) illustrates this fact for graphene on five different dielectrics with increasing low-frequency dielectric constants at room temperature for an impurity density of $n_{imp} = 5 \times 10^{11}$/cm$^2$ and a carrier concentration $n = 10^{12}$/cm$^{2}$.  The mobility ($\mu=\sigma/ne$) is calculated from the conductivity using the relation $\sigma(T)=(e^{2}/h)2v_{F}k_{F}\tau$.  The solid curve is the predicted enhancement of mobility due to the damping of Coulombic impurity scattering, and the hollow circles indicate the expected mobility for the particular dielectrics if SO phonon scattering was absent (these are calculated with $\tau = \tau_{imp}$).  The filled circles show the degradation of the mobility due to SO phonon scattering; for these values of mobility, $\tau_{tot} = ( \tau_{ph}^{-1} + \tau_{imp}^{-1} )^{-1}$ is used.  It is evident that SO phonon scattering nearly washes out the improvement of mobility that could have resulted from the reduction of impurity scattering, and RT mobilities are limited to $\mu_{RT} \sim 10^{4}$ cm$^2$/V$\cdot$s, as indicated by the shaded band.  The effect of SO phonon scattering might seem minimal for SiO$_{2}$, but it is important to note that the SO phonon scattering is independent of the impurity concentration; therefore a purer graphene/oxide interface will still result in substantial SO phonon scattering and mobilities in this regime.  In Fig \ref{Fig4}(b) we plot the carrier conductivity in graphene for two different materials, SiO$_{2}$ and ZrO$_{2}$.  The solid lines correspond to conductivity due to impurity scattering alone, while the dashed lines incorporate the effect of both impurity and surface-optical phonons.  The conductivity is linear with carrier concentration in agreement with experiments \cite{ChenNatNano08}. Due to 
\begin{figure}[h]
\includegraphics[height=65mm, width=85mm]{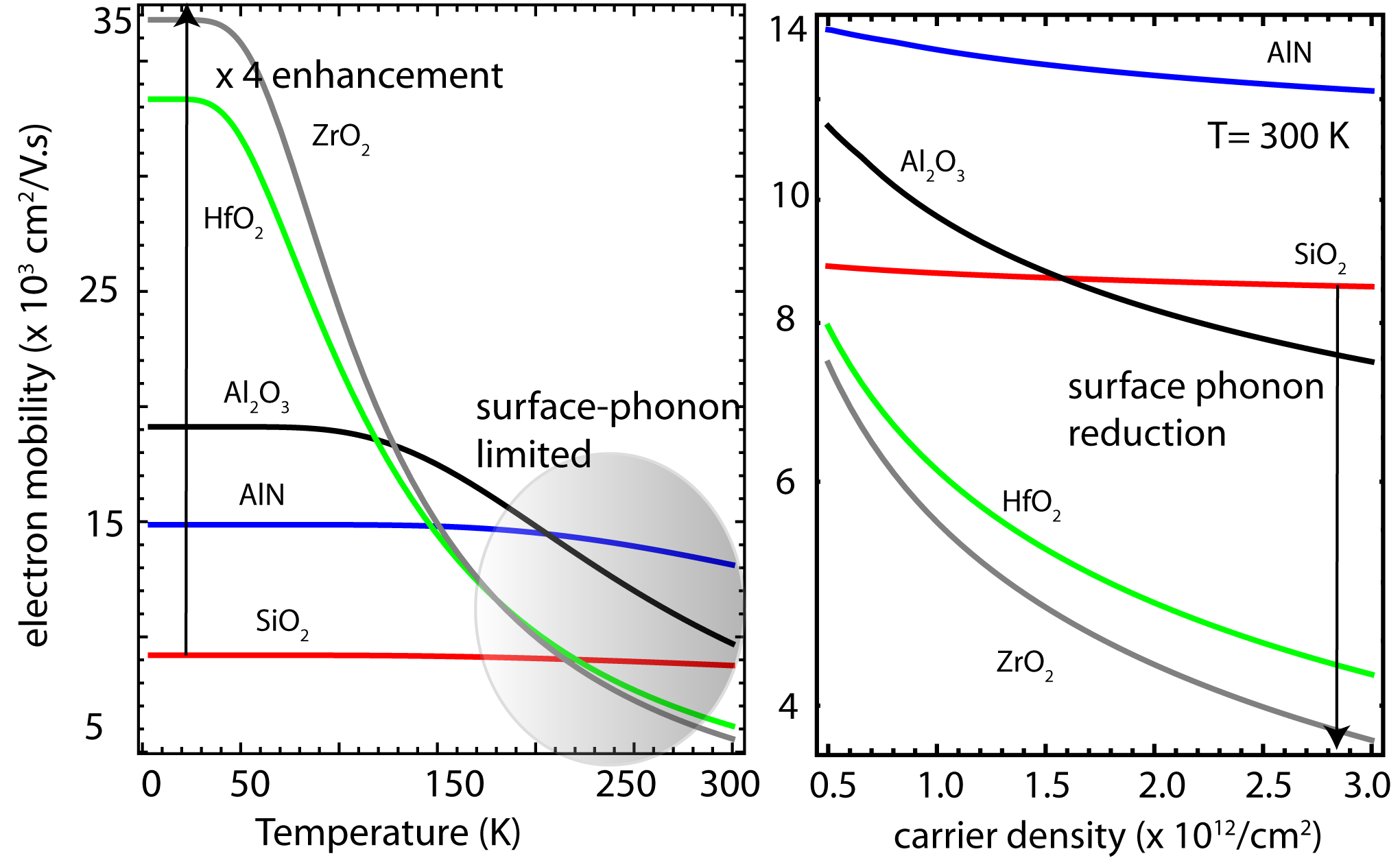}
\caption{ (Left - (a): Electron mobility in graphene as a function of temperature on different gate dielectrics at carrier density n= 10$^{12}$/cm$^{2}$.  Note that high-$\kappa$ dielectrics reduce Coulombic impurity scattering and increases carrier mobility by a factor of 4 at low temperatures, but strong SO phonon scattering by them ( shaded part ) reduces the RT mobilities to $\sim 10^{4}$ cm$^2$/V$\cdot$s.  Right - (b): Electron and hole conductivity as a function of carrier concentration for graphene on different dielectrics at T=300K. For all calculations above impurity density is assumed to be $5\times 10^{11}$/cm$^2$.}
\label{Fig5_1}
\end{figure} 
the importance of SO phonon scattering, mobility is also expected to be temperature-dependent Ñ with lowering of temperature, the reduction of SO phonon scattering will result in an increase in mobility from the filled circles in Fig \ref{Fig4}(a) to the hollow circles. Figure \ref{Fig5_1}(a) shows the temperature dependent carrier mobility for graphene on different dielectrics. Indeed, at low temperatures, electron mobility can be improved by a factor of 4 (gray curve in Fig \ref{Fig5_1}) by putting graphene on high-$\kappa$ dielectrics (here ZrO$_{2}$) but at room temperatures, mobility values are close to $10^{4}$ cm$^2$/V$\cdot$s on all dielectrics due to strong surface-phonon scattering. The carrier concentration (n) dependence of carrier mobility can be understood by simple mathematical arguments. As square of scattering matrix element $ |\mathcal{M}_{q}^{\nu}(q)|^{2} \sim1/q$ (see eq.\ref{matrix}) and graphene density of states (DOS) $g(q)\sim q$ \cite{TFangAPL07}, without screening, scattering rates $\tau^{-1}\sim|\mathcal{M}_{q}^{\nu}(q)|^{2} \times g(q)$ ={\it{constant}} (saturation behavior of scattering rates as shown in Fig.\ref{Fig3}(a)). Hence, conductivity $\sigma(n)\sim k_{F}$ and electron mobility, without screening monotonically decreases as $\mu(n)\sim1/\sqrt{n}$. This monotonic decrease of mobility with electron density is shown in Fig. \ref{Fig5_1}(b). Numerical fittings show that mobility varies with carrier density following a power law behavior $\mu(n)\sim$n$^{-\alpha}$, where the exponent $\alpha$ lies between 0.3-0.4. The minor deviation of $\alpha$ from 1/2 \cite{FratiniPRB08} is due to inclusion of Thomas-Fermi screening in scattering rate calculations. For accurate prediction of the exponent $\alpha$, a sophisticated  dynamic screening function ($\epsilon(q,\omega)$) is needed because Thomas-Fermi screening in the static limit ($\epsilon(q,0)$) is inappropriate for a dynamic perturbation like phonon vibration. It is worthwhile at this point to discuss some recent experimental results of graphene transport in high-$\kappa$ environments. Experiments reported contradictory results for graphene submerged in liquid high-$\kappa$ environments, whereas results for graphene on crystalline substrate are in accordance with our theoretical prediction. Details of these experiments are discussed in appendix A.  \\
Among the dielectrics considered here, AlN is the most promising for maintaining relatively high mobilities due to the higher SO phonon frequencies compared to Al$_2$O$_3$, HfO$_2$ and ZrO$_2$ resulting in lower SO phonon scattering, and a higher static dielectric constant than SiO$_2$ resulting in lower impurity scattering.  Along the same lines, epitaxial single-layer graphene grown on SiC substrates are expected to be less affected by SO phonon scattering due to the high phonon energies of crystalline SiC (see Table \ref{table1}, also pointed out in \cite{FratiniPRB08}), and at the same time have a lower sensitivity to impurity scattering due to the higher dielectric constant  of SiC compared to SiO$_2$. If one compares carrier mobilities in graphene with 2DEGs in III-V semiconductor heterostructures, the impurity scattering component in InAs or InSb based heterostructures is strongly damped by modulation-doping, which allows the spatial separation of charged impurities from mobile carriers.  At the same time, narrow-bandgap semiconductors have intrinsically high static dielectric constants ($\kappa_{InAs} \sim 12.5$ and $\kappa_{InSb} \sim 18$).  Therefore, impurity scattering can be effectively damped in such materials, and the RT mobilities are typically limited by interface roughness scattering (for thin wells), or polar optical phonon scattering, which is strong in III-V semiconductors. The fact that graphene is an atomically thin layer makes its transport properties especially sensitive to the surrounding dielectrics, and by judicious choice of such dielectrics, high low-field charge mobilities can be retained (nominal enhancement) as well as better capacitive charge control can be achieved. The ideal dielectrics would be those that possess both high static dielectric constants, and high phonon energies that are not activated in low-field transport.  Among the dielectric materials considered, SO phonon scattering from them limits the mobilities to lower values than corresponding III-V semiconductors such as InAs and InSb.  Suspending graphene removes the SO phonon scattering component, but reduces the electrostatic control by a capacitive gate and introduces electro-mechanical effects.  While this can be a feasible route towards ballistic transport in graphene, a high-$\kappa$ dielectric with high SO phonon energies would be the most desirable in the future.
\section{conclusions}
In conclusion, we have investigated the effect of various dielectrics on the electron mobility in single layer graphene by considering the effects of impurity and SO phonon scattering.  By calculating the remote polar optical phonon scattering arising from the polar substrates, and combining it with their effect on Coulombic impurity scattering, a comprehensive picture of the effect of dielectrics on charge transport in graphene emerges.  We have shown that at low temperatures, high-$\kappa$ dielectric reduces the charged impurity scattering resulting 4-5 times enhancement of electron mobility in graphene in addition to improved capacitive charge control of the g-FET . At room temperatures this mobility enhancement is washed out (mobility is reduced compare to graphene on low-$\kappa$ dielectrics such as SiO$_{2}$) by strong surface-optical phonon scattering leaving little or no room for mobility improvement.  Among commonly available dielectrics, intermediate-$\kappa$ dielectrics (such as AlN and SiC) are optimum choice as gate-insulators where nominal mobility enhancement is possible ($\mu \sim 15,000$ cm$^{2}$/V.s) compare to ubiquitously used low-$\kappa$ gate-dielectric SiO$_{2}$.
\section{Acknowledgement}
The authors would like to acknowledge  S. Basuray, S. Senapati (University of Notre Dame), V. Perebeinos ( IBM T. J. Watson) for helpful discussions and National Science Foundation (NSF), Midwest Institute for Nanoelectronics Discovery (MIND) for the financial support for this work.\appendix
\section{Recent Experimental results}
 In this section we will discuss some of the recent transport experiments for graphene based devices in different dielectric media. To our best knowledge, we are aware of three experimental reports so far for graphene in high-$\kappa$ environments. Among them, two experiments \cite{ponomarenkoPRL, ChenNanoletter09} were done using high-$\kappa$ organics solvents and reported contradictory results.  As our theory is restricted to crystalline gate-oxides, a quantitative comparison of experiments with liquid gates to our theory is beyond the scope of the work presented here.  Nevertheless, a qualitative explanation can be made for the sake of completeness of charge transport in high-$\kappa$ dielectrics. Ponomarenko et. al \cite{ponomarenkoPRL} has reported graphene mobility to be insensitive to dielectric environments while Chen et. al \cite{ChenNanoletter09} reported order of magnitude enhancement in carrier mobility. The insensitivity of carrier mobility on environmental dielectrics as reported in ref. 17 can well originate from two reasons: Ñi) The gate-leakage current due to insufficient electrical isolation can results in erroneous measurement. ii) Moreover, formation of Stern layer \cite{KrugerAPL01} at graphene-liquid interface will increase charged impurity scattering and compensate the effect of dielectric screening partially, if not by full extent.  On the other hand, Chen et al. performed experiments with proper electrical isolation (covered electrodes with Ti) and used SiO$_{2}$ back gate instead of liquid top gate. Use of high-$\kappa$ aprotic (no hydrogen bonds) organic solvent not only enhanced Coulomb screening but also compensated charged impurity by inducing counter ions at graphene-liquid interface \cite{Chennanoletter09}. Long range surface-phonon vibrations are absent in liquids and surface-phonon of crystalline SiO$_{2}$ backgate are not activated low-field transport at room temperatures (see red curve of Fig. 5). As a result, Chen et al. has observed a significant increase of carrier mobility using high-$\kappa$ aprotic liquids. \\
 For practical graphene-based field effect devices (g-FET), use of liquid environments is not a feasible idea and a solid (crystalline) high-$\kappa$ environment is desirable for large-scale production.  The 3$^{rd}$ and most recent experiment \cite{ZouPRL} has reported transport results of graphene-FET using HfO$_{2}$ as top gate.  Though high-$\kappa$  HfO$_{2}$ top gate enhances electrostatic doping in graphene, strong surface-phonon field arising from top gate together with impurity scattering limits the carrier mobility around 10,000 cm$^{2}$/V.s. { \it{This is a direct experimental evidence of the theoretical work presented in this paper}}. It should be noted that  experimentally extracted value of HfO$_{2}$ SO-phonon limited mobility ($\sim20,000$cm$^{2}$/V.s) is higher than our calculated value ($\sim15,000$cm$^{2}$/V.s).  This discrepancy stems out form the fact that in the experiment by Zou et al. thin ($\sim $ 10 nm) HfO$_{2}$ top gate was amorphous in nature with a single SO-phonon mode at 54 mev,  whereas in our theory, thick HfO$_{2}$ crystalline gate has a low-lying SO-mode ($\sim $ 20 mev, see TableI) in addition to the high-energy mode. The exclusion of low-energy SO mode from our scattering calculations will result in SO-phonon limited mobility reasonably close to the experimentally obtained  value. 
\section{Details of electron-SO phonon scattering strength }
Here we will outline the detail calculation of electron-SO phonon interaction potential. 
The potentials in Eq.\ref{potential} are undetermined by a constant $b_{q,\omega}$ and $d_{q,\omega}$ is related to $b_{q,\omega}$ by the relation $d_{q,\omega}=2b_{q,\omega}\sinh(qt_{ox})$. This constant can be determined by semiclassical argument originally proposed by Stern and Ferrel \cite{Stern} and later employed by Fischetti for 2D nanoscale MOSFETS \cite{FischettiJAP01}. We first consider the time-averaged total electrostatic energy, $\langle\mathcal{W}^{(i)}_{q}\rangle$  (bra-kets $\langle....\rangle$) stands for time average) associated with the field $\phi_{q}^{(i)}({\bf{r}},t)$ caused by the excitation of mode $i$ oscillating at the frequency $\omega^{(i)}_{q}$ (root of the secular equation \ref{secularequation}). As  fields are harmonic oscillations in time, the total energy is just the twice of the time-averaged potential energy,  $\langle\mathcal{W}^{(i)}_{q}\rangle=2\langle\mathcal{U}^{(i)}_{q}\rangle$.  By evaluating electric filed as ${\bf{E}}^{i}({\bf{r}},t)=-\nabla\phi_{q}^{(i)}({\bf{r}},t)$, the electrostatic energy can be expressed as
\begin{eqnarray}
\langle\mathcal{W}^{(i)}_{q}\rangle&=&2\langle\mathcal{U}^{(i)}_{q}\rangle\nonumber\\
&=&2\Big\langle\int_{-\infty}^{\infty}dz\int_{0}^{\infty}rdr\epsilon(\omega^{(i)}_{q},r)|{\bf{E}}^{i}({\bf{r}},t)|^{2}\Big\rangle\nonumber\\
&=&d_{q,\omega}Aq\Big[\epsilon_{ox}(\omega)\coth(qt_{ox})+1\Big].
\end{eqnarray}
The unknown constant $d_{q,\omega}$ now can be determined by equating the total electrostatic energy to the energy of quantized harmonic oscillator in their ground state, i.e $\langle\mathcal{W}^{(i)}_{q}\rangle = \hbar\omega^{(i)}_{q}/2$. Furthermore, polarization field associated with $\nu^{th}$ SO-phonon is given by the difference of square amplitudes of high frequency ($\epsilon_{ox}(\omega)=\epsilon_{ox}^{\infty}$) and zero-frequency ($\epsilon_{ox}(\omega)=\epsilon_{ox}^{0}$) response of bulk insulator phonons as outlined by Fischetti et. al \cite{FischettiJAP01, comment}.  Implementing the procedure outlined above, the scattering potential inside graphene membrane for $\nu^{th}$ SO-phonon can be written as (in the limit qt$_{ox}\rightarrow\infty$)
\begin{eqnarray}
{\mathcal{H}}_{in}=&&\Big[\frac{e\hbar\omega_{SO}^{\nu}}{2Aq}\Big(\frac{1}{\epsilon_{ox}^{\infty}+1}-\frac{1}{\epsilon_{ox}^{0}+1}\Big)\Big]^{\frac{1}{2}}\exp(-qz) \nonumber\\
&&\times(e^{i{\bf{q}}.{\bf{r}}}a^{\nu\dagger}_{q}+e^{-i{\bf{q}}.{\bf{r}}}a^{\nu}_{q}).
\label{amplitude}
\end{eqnarray}
Comparing with eq.\ref{interaction}, we get the expression of ${\mathcal{F}}_{\nu}$ as given in the eq.\ref{F}.


\begin{thebibliography}{99}
\bibitem{comment} If there exists more than one TO phonon modes in the insulator, it is non-trivial to separate-out individual contributions of different TO-phonon modes to a single SO-phonon mode. Section C of ref.19 illustrates a systematic procedure to separate-out contributions of different TO modes on a single SO mode. In our case, we assumed oxides have two bulk TO modes and they are weakly coupled together via eq.\ref{secularequation}. Nevertheless, we can ignore this weak coupling between two TO-phonons (because of their large difference in energy) and assume SO1 (SO2) mode is arising solely due to TO1 (TO2) phonon without incorporating significant error in the calculations.
\end{thebibliography}
\end{document}